\documentclass[letterpaper, 10 pt, conference]{ieeeconf}
\IEEEoverridecommandlockouts
\usepackage{cite}
\usepackage{amsmath,amssymb,amsfonts,mathtools}
\usepackage{amsthm}
\usepackage{algorithmic}
\usepackage{graphicx}
\usepackage{float}
\usepackage{textcomp}
\usepackage{xcolor}

\newtheorem{problem}{Problem}

\newtheorem{proposition}{Proposition}
\usepackage{relsize}
\usepackage{bm}
\usepackage{soul}
\usepackage{cancel}
\usepackage[hidelinks]{hyperref}
\newcommand{\fref}[1]{Fig.~\ref{#1}}

\def\BibTeX{{\rm B\kern-.05em{\sc i\kern-.025em b}\kern-.08em
    T\kern-.1667em\lower.7ex\hbox{E}\kern-.125emX}}
\begin{document}

\title{Modeling Adaptive Tracking of Predictable Stimuli in Electric Fish 
\thanks{This work was supported by the Office of Naval Research, United States under grant N00014-21-1-2431.}
\thanks{$\dagger$ These authors contributed equally to this work.}
\thanks{$^1$ John A. Paulson School of Engineering and Applied Sciences, Harvard University, Allston, MA, 02134, USA; yangyu@seas.harvard.edu.}
       \thanks{$^2$ Department of Electrical and Computer Engineering, Northeastern
University, Boston, MA, 02115, USA; franciscodemelooli.a@northeastern.edu, msznaier@coe.neu.edu. }
\thanks{$^3$ Department of Mechanical Engineering, Johns Hopkins University, Baltimore, MD, 21218, USA; llw@jhu.edu, ncowan@jhu.edu} 
\thanks{$^4$ Department of Electrical Engineering, Escola Politécnica da Universidade de São Paulo, Sao Paulo, SP, 05508-010, Brazil; pait@usp.br.}
}

\author{Yu Yang$^{1,3\dagger}$ 
\and
Andreas Oliveira$^{2\dagger}$
\and
Louis L. Whitcomb$^3$
\and
Felipe Pait$^4$
\and
Mario Sznaier$^2$
\and 
Noah J. Cowan$^3$
}

\maketitle

\begin{abstract}
The weakly electric
  fish \emph{Eigenmannia virescens} naturally swims back and forth to
  stay within a moving refuge, tracking its motion using visual and
  electrosensory feedback. Previous experiments show that when the
  refuge oscillates as a low-frequency sinusoid (below about 0.5 Hz),
  the tracking is nearly perfect, but phase lag increases and gain
  decreases at higher frequencies. 
  Here, we model this nonlinear behavior as an adaptive internal model principle
  (IMP) system. Specifically, an adaptive state estimator
  identifies the \emph{a priori} unknown frequency, and feeds this
  parameter estimate into a closed-loop IMP-based system built
  around a lightly damped harmonic oscillator.  We prove that the closed-loop
  tracking error of the IMP-based system, where the online
  adaptive frequency estimate is used as a surrogate for the unknown
  frequency, converges exponentially to that of an ideal control
  system with perfect information about the stimulus. Simulations
  further show that our model reproduces the fish refuge tracking Bode
  plot across a wide frequency range. These results establish the
  theoretical validity of combining the IMP with an adaptive
  identification process and provide a basic framework in adaptive
  sensorimotor control.
\end{abstract}

%\begin{keywords}
%Internal Model Principle, Adaptive Systems Theory, %Bio-Inspired Control.
%\end{keywords}

\section{Introduction}

\begin{figure}[h]
    \centering
\includegraphics[width=0.9\linewidth]{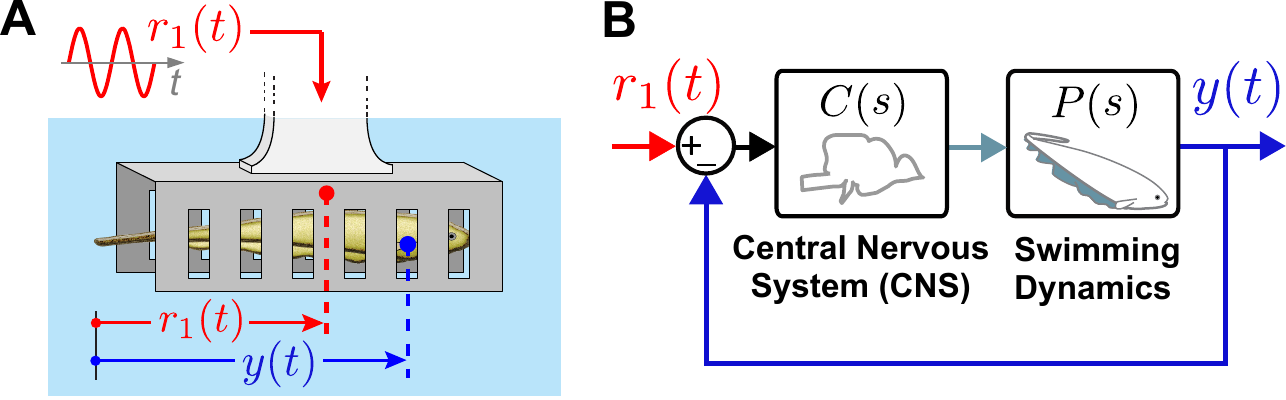}
    \caption{(A) Weakly electric fish tracks a one-degree-of-freedom moving refuge controlled by single sinusoidal reference stimulus $r_1(t)$ and the fish position is $y(t)$. (B) Fish control their movement to track the moving refuge through feedback, sending the error signals to the controller central nervous system (CNS) and the plant swimming dynamics.}
    \label{fig1}
\end{figure}

Understanding how animals integrate sensory information to guide
locomotion is an important field of study in biology, control
engineering, and robotics
\cite{aguilar2016review,madhav2020synergy,rothcomparative2014}. One
such animal, the weakly electric glass knifefish (\emph{Eigenmannia
  virescens}), is an ideal model system for studying sensorimotor
control. These fish have two independent image-forming
senses (vision and electrosense)
\cite{cowan2014feedback,stamper2012active,sutton2016dynamic,yeh2024illumination}
and naturally swim back and forth to maintain their body position
within a moving polyvinyl chloride (PVC) refuge
\cite{cowan2007critical,roth2011stimulus,yang2020comparison,yang2024sensorimotor}
(\fref{fig1}A). They effectuate this sensorimotor tracking behavior by
modulating the undulatory dynamics of an elongated fin on their
ventral side (\fref{fig1}B), a locomotor mechanism that is
accurately modeled by a mass--damper system
\cite{sefati2013mutually,uyanik2020variability}.

Previous research on \emph{Eigenmannia} examining tracking performance
for sinusoidal stimuli in the frequency range of 0.10 Hz to 2.05 Hz
show that the fish track nearly perfectly at the lower end of this
frequency range \cite{roth2011stimulus}, i.e.\ a closed-loop gain near
1 with minimal phase lag. However, tracking performance degrades as
the stimulus frequency increases. Moreover, fish exhibit a fundamental
nonlinearity: single-sine responses do not predict sum-of-sine
responses via linear superposition. Specifically, when a low-frequency
sinusoid is presented in isolation, fish track with substantially
lower phase lag and better gain than when that signal is included
within a complex, multi-frequency signal (i.e.,\ sum-of-sines)
\cite{roth2011stimulus}. This nonlinear behavior has also been
observed in humans
\cite{zimmet2020cerebellar,wilkinson2025active}. One hypothesis is
that single sinusoidal signals are ``predictable'', therefore fish
(and humans) can build an internal model for them
\cite{roth2011stimulus}.

This paper presents a novel control-theoretic model on how fish
adaptively build an internal model for the external single sinusoidal
stimulus. The model includes two components. First, from the
literature
\cite{francis1976internal,gillespie2016human,huang2018internal,cutlip2019haptic},
the internal model principle (IMP) shows that a controller having a
model of input stimulus can perfectly track that signal at
steady-state just using moderate loop gains. More specifically, it has
been shown that a sinusoidal input can be perfectly tracked in
steady-state given that a harmonic oscillator with poles at the input
frequency is coupled into the feedback controller
\cite{huang2018internal,cutlip2019haptic}. Given the experimental
evidence that fish \emph{nearly} perfectly track the low-frequency
single sinusoidal moving refuge but the tracking degrades at higher
frequencies, we model the fish refuge tracking system by coupling a
\emph{lightly damped} harmonic oscillator into its controller. Second,
since the damped harmonic oscillator contains the stimulus frequency
which is not directly known by the fish, in our model, a nonlinear
time-varying adaptive identifier \cite{narendra1989stable} is created
to continuously estimate the frequency of the external sinusoidal
signal in real time, modeling the process by which fish adaptively
identify the stimulus frequency. Taken together, this control model
takes advantage of two concepts that are well-known yet present
significant challenges when combined, i.e., the IMP and an adaptive
(frequency) identification process.

To the best of our knowledge, the most closely related research
directions are the certainty equivalence principle
\cite{Morse-Certainty-Equivalence} and adaptive internal model control
\cite{Ochoa-IMP-Adaptive}. The certainty equivalence principle
concerns designing a stabilizing controller based on current estimates
of plant parameters produced by an identification process. In
\cite{Morse-Certainty-Equivalence}, it is shown that even if the plant
parameters fail to converge, controllers designed under this principle
remain justified in the sense that the family of parametrized
closed-loop systems is detectable. This result, though, provides no
guarantee that the tracking error of a plant relying on adaptive
estimates converges to that of a plant with perfect information. The
work in \cite{Ochoa-IMP-Adaptive} addresses this gap by showing that
such convergence occurs, but only through an asymptotic argument
without an explicit convergence rate. Moreover, their analysis is
restricted to stable open-loop plants, which does not encompass the
setting considered here, since we are interested in frequency
estimation of a sinusoidal signal.

In this paper, we establish the theoretical validity of designing a
controller based on adaptive estimates of an external sinusoidal
stimulus. First, we show that in our adaptive identifier, both the
state and frequency estimates converge globally to their true values
at an exponential rate; this analysis assumes that the input stimulus
is a sinusoid (of unknown frequency and phase). Second, we show that
the states of a closed-loop IMP-based system---operating with a
real-time estimate of the stimulus frequency fed in from the adaptive
estimator---converges globally exponentially to the states of an
idealized and stable closed-loop system that assumes perfect access to
the true resonant frequency.  In addition to building on the prior
theoretical work described above
\cite{Ochoa-IMP-Adaptive,Morse-Certainty-Equivalence}, we test the
experimental predictions of our model, successfully capturing the
frequency response plot of \emph{Eigenmannia} across a wide frequency
range using data from three fish. Thus, while one modeling study
cannot establish that this is the precise neural mechanism animals
use, it nevertheless provides a compelling and rigorous modeling
framework for analyzing adaptive target tracking behavior both
theoretically and experimentally \cite{mongeau2024moving}, fostering
comparative studies across taxa.

%The paper is organized as follows: Section II motivates and formulates the model used throughout the paper, along with the problem definitions. Section III presents a proof for the exponential convergence of the adaptive identifier. Section IV presents a proof for the exponential convergence of the internal model principle based controller using internal estimates. Section V demonstrates the model's ability to explain fish tracking data. Finally, Section VI presents conclusions and directions for future research.

\section{Model}

\subsection{Motivation}
\label{sec:motivation}
Experimental results in the literature \cite{cowan2007critical,roth2011stimulus,stamper2012active} show that fish track single sinusoids almost perfectly at low frequencies (below about 0.5 Hz), suggesting that such inputs are predictable and fish can build an internal model for them. Motivated by this observation, we invoke the IMP to model the refuge tracking system with a harmonic oscillator coupled inside the controller (\fref{fig2}).
 
If the tracking were perfect across the entire frequency bandwidth, it
would be natural to conclude that the fish's controller implements a
perfect harmonic oscillator as dictated by the IMP. However,
experimental evidence \cite{roth2011stimulus} shows that increasing
the stimulus frequency leads to degraded tracking performance, with
phase lag increasing and gain decreasing. To account for this
behavior, we propose that fish instead implement a \emph{damped
  harmonic oscillator} in the controller, as detailed below. Lastly,
since the position, velocity and frequency of oscillation of the
refuge are unknown to the fish, we hypothesize that it utilizes an
adaptive identification scheme to continuously estimate these
quantities during experiments and update the resonant frequency in the
damped harmonic oscillator. Biologically, these computations likely occur in cerebellar-like structures 
 \cite{miall2008state,haar2020revised,krakauer2011human}.

% We show later in the formulation below that the tracking error when the fish is correctly estimating the resonant frequency is proportional to $\zeta \omega_0^2$, where $\zeta$ is the damping ratio, which agrees with experimental results. 

% Thus far the above model can be fully done via linear system analysis in fact that is why we can obtain the tracking error.  is required to validate the model qualityis since the fish does not have a perfect estimate of the position of the PVC and the frequency of its oscillation, we assume it utilizes an adaptive identifier scheme, to derive these two quantities. 

\subsection{Formulation}
The refuge
position, $r_1(t)$, is sinusoidal  with unknown frequency $\omega_0$ and phase $\phi$. Denoting the refuge velocity by $r_2(t) = \dot{r}_1(t)$, we have
\begin{align}
\label{eq:functiongenerator}
\dot{r}(t)=\begin{bmatrix}
\dot{r}_1(t)\\\dot{r}_2(t) 
\end{bmatrix}
=\begin{bmatrix}
0 & 1\\-\omega_0^2 & 0 
\end{bmatrix}r(t).
\end{align}
Denoting $\theta = -{\omega}^2_0$ as the unknown frequency parameter, \eqref{eq:functiongenerator} can be written as 
\begin{align}
\begin{split}
\dot{r}(t)
&= \begin{bmatrix}
0 & 1\\0 & 0 
\end{bmatrix}r(t)+ \theta\begin{bmatrix}
0 & 0\\1 & 0 
\end{bmatrix}r(t),\\
&= M^\top r(t)+\theta M r(t).
\end{split}
\label{eq:r_equation}
\end{align}
%
% Here $\theta$ is the unknown parameter related to the frequency of single sine stimulus that needs to be identified.
We model the adaptive identifier implemented by the fish as
\begin{multline}
  \label{eq:identifier}
\dot{\hat{r}}(t)
= M^\top \hat{r}(t)+\hat{\theta}(t)Mr(t) +  (A_m - M^\top)[\hat{r}(t)-r(t)],
\end{multline}
where $\hat{r}(t) = [\hat{r}_1(t) \ \hat{r}_2(t)]^\top$ is the estimated input stimulus, $\hat{\theta}(t)$ is the estimated frequency parameter with estimated  frequency $\hat{\omega}(t) = \sqrt{-\hat{\theta}(t)}$, and $A_m \in \mathbb{R}^{2\times2}$ is a Hurwitz matrix, i.e., all eigenvalues of $A_m$ have negative real part. Note that the adaptive identifier assumes full state measurement of the moving stimulus, $r(t)$. The parameter update law for $\hat \theta(t)$ takes the form
\begin{equation}
  \dot{\hat{\theta}}(t) =-\gamma \Delta r^\top(t) P M r(t),
  \label{eq:updatelaw}
\end{equation}
where $\gamma>0$ is the gain of the update law (also known as the ``adaptive gain'') and $P\in\mathbb{R}^{2\times2}$ is a yet unspecified positive definite matrix. Given the design of the adaptive identifier, the first problem is posed as follows:
\begin{problem}\label{prob1}
    Does $\lim_{t \rightarrow \infty} \hat{r}(t) \rightarrow r(t)$ and $\lim_{t \rightarrow \infty} \hat{\omega}(t) \rightarrow \omega_0$? If so, is the convergence exponential?
\end{problem}
\begin{figure}[H]
    \centering
\includegraphics[width=\linewidth]{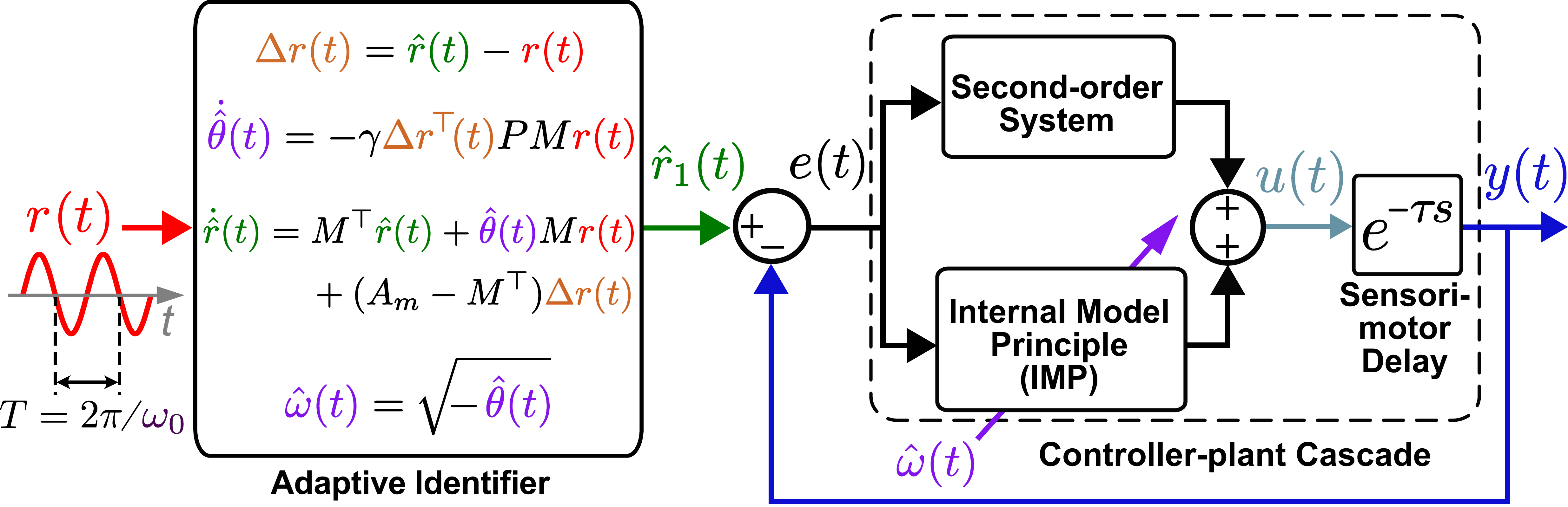}
 %   \caption{A block diagram that illustrates the model of fish tracking a single sine stimulus with frequency $\omega_0$. The stimulus $r(t) = [r_1(t) \ r_2(t)]^\top$ contains the refuge position $r_1(t)$ and refuge velocity $r_2(t)$ and is passed through an adaptive identifier. The difference between the adaptively identified refuge position $\hat{r}_1(t)$ and the fish position $y(t)$ is the sensory error $e(t)$ that passes through a controller $C$, a plant $P(s)$ and a sensorimotor delay $e^{-\tau s}$. The controller $C$ is composed of a proportional-derivative (PD) controller and an IMP-based controller with a damped harmonic oscillator coupled inside. The identified frequency $\hat{\omega}(t)$ keeps updating the parameters in the damped harmonic oscillator, thus the controller itself is time-varying.}
        \caption{A block diagram that illustrates the model of fish tracking a single sine stimulus with frequency $\omega_0$. The stimulus $r(t) = [r_1(t) \ r_2(t)]^\top$ contains the refuge position $r_1(t)$ and refuge velocity $r_2(t)$ and is passed through an adaptive identifier. The difference between the adaptively identified refuge position $\hat{r}_1(t)$ and the fish position $y(t)$ is the sensory error $e(t)$ that passes through the controller-plant cascade with a sensorimotor delay $e^{-\tau s}$. The controller-plant cascade contains a second-order system and an IMP pathway that is a damped harmonic oscillator connected in parallel. The identified frequency $\hat{\omega}(t)$ keeps updating the parameters in the damped harmonic oscillator, thus the system is time-varying.}
    \label{fig2}
\end{figure}
As described in Section~\ref{sec:motivation}, we hypothesize that fish
adaptively identify the refuge motion $\hat{r}(t)$ and the frequency
$\hat{\omega}(t)$ from $\hat{\theta}(t)$ in their closed-loop system
(\fref{fig2}). The sensory error $e(t) = \hat{r}_1(t) - y(t)$ is
determined by the fish position $y(t)$ and the state estimate of
refuge position $\hat{r}_1(t)$. Note that one could assume the fish
uses the ``raw'' sensory measurement, $r_1(t)$, but a normative
engineering approach (and the one we take here) is to use an estimator
that ``smooths'' the input.

%Based on experimental data, we model the refuge tracking system controller using 1) a proportional-derivative (PD) controller and 2) an IMP-based controller that contains a \emph{damped harmonic oscillator} with a time-varying frequency $\hat{\omega}(t)$ connected in parallel (\fref{fig2}). Therefore, the controller $C$ is a time-varying system. When $\hat{\omega}(t)$ converges to the stimulus frequency $\omega_0$ in steady state, which will be shown later in this paper, the steady-state controller, plant $P(s)$, and sensorimotor delay $e^{-\tau s}$ cascade becomes time-invariant and can be represented using a transfer function of the form

Based on experimental data \cite{yangdissertation2025}, we model the refuge tracking system controller-plant cascade using 1) a second-order system and 2) an IMP that is a \emph{damped harmonic oscillator} with a time-varying frequency $\hat{\omega}(t)$ that are connected in parallel and passed through a sensorimotor delay (\fref{fig2}). Therefore, the system is a time-varying system. When $\hat{\omega}(t)$ converges to the stimulus frequency $\omega_0$ in steady state, which will be shown later in this paper, the steady-state controller-plant cascade becomes time-invariant and can be represented using a transfer function of the form
%
%\begin{equation}
%\begin{aligned}
%\bigg(\underbrace{k_d s + k_p}_\mathrm{PD  \ %controller} + \underbrace{\frac{k_\mathrm{IMP}(s^2 %+k_{\mathrm{plant}}s)}{s^2 + 2\zeta{\omega
%_0}s + \omega_0^2}}_{\mathrm{IMP-based  \ controller}} %\bigg)\underbrace{\frac{\kappa}{s^2 + %k_{\mathrm{plant}} s}}_{\mathrm{Plant}} \ \ e^{-\tau s}.
%\end{aligned}
%\label{eq:tf}
%\end{equation}
%
%
\begin{equation}
\begin{aligned}
\bigg( \underbrace{\frac{k_1 s + k_2}{s^2 + k_3 s}}_{\mathrm{2^{nd} \ order \ system}} + \underbrace{\frac{k_4}{s^2 + 2\zeta\omega_0 s + \omega_0^2}}_{\mathrm{Damped  \ oscillator}} \bigg) e^{-\tau s},
\end{aligned}
\label{eq:tf}
\end{equation}
% 
%Note that the denominator of the IMP-based controller is a second-order polynomial that has a damping coefficient $\zeta$ in addition to the denominator of a perfect harmonic oscillator. Furthermore, in our model structure, we connect two parts of the controller in parallel (\fref{fig2}) rather than the traditional way in series (\fref{fig1}C), but it is easy to show that a damped harmonic oscillator can be factored out from the transfer function of the steady-state controller, plant, and sensorimotor delay cascade which is similar to the block diagram in \fref{fig1}C. Also, by a simple re-factorization, the transfer function
%
where $\zeta$ is the damping coefficient in the damped harmonic
oscillator. In reality, we must assume the fish does not know
$\omega_0$, and design a controller based only on estimated frequency
and state values; thus we ask the following:
%\begin{equation}
%\begin{aligned}
%\bigg(\frac{k_1 s + k_2}{s^2 + k_3 s} + \frac{k_4}{s^2 + 2\zeta\omega_0 s + \omega_0^2} \bigg) e^{-\tau s} 
%\\ = ,
%\end{aligned}
%\label{eq:openloop}
%\end{equation}
% 
% where $k_1 = k_d \kappa$, $k_2 = k_p \kappa$, $k_3 = k_{\mathrm{plant}}$, $k_4 = k_{\mathrm{IMP}} \kappa$. 
% The system \eqref{eq:tf} is later used for our proof of the second problem as follows:

%As demonstrated in the Experimental Section (Figures \AO{Yu Yang Insert Figures}), when $\hat{\omega}(t) = \omega_0$ and $\hat{r}_1 = r_1$, the proposed model accurately reproduces the phase lag and gain exhibited by the fish refuge tracking mechanism. To ensure that this model remains meaningful when only estimated frequency and state values are available, it is therefore natural to require that these estimates converge to their true counterparts in the limit. Motivated by this, we pose the following problems:

\begin{problem}\label{prob2}
  Consider the idealized sensory error $e_c(t)$ defined assuming
  $\omega_0$ and $r_1(t)$ were known (instead of their estimates as in
  \fref{fig2}). Is the error system defined by the discrepancy between
  the actual and idealized sensory errors, namely $e(t) - e_c(t)$,
  globally exponentially stable?
\end{problem}

%Note that if the statements in the second problem are true, then it is indeed sensible for the fish to use internal estimates of frequency and position to implement its internal model controller.

\section{Adaptive Identification}
\label{sec:adaptive_model}

Our objective is to show that the adaptive identifier is asymptotically stable and that the state estimate and parameters converge thereby answering \textbf{Problem }\ref{prob1}. We define error coordinates as follows:
\begin{equation}
    \Delta r(t) = \hat{r}(t)-r(t),
\end{equation}
\begin{equation}
\Delta \theta(t) = \hat{\theta}(t)-\theta.
\end{equation}
Substituting in \eqref{eq:r_equation}, \eqref{eq:identifier}, and
\eqref{eq:updatelaw}, this error system can be written as:
\begin{align}
  \begin{split}
   \Delta \dot{r}(t) 
                    % &= \dot{\hat{r}}(t)-\dot{r}(t)
                     % &= M^\top \Delta r(t) + \Delta \theta(t)Mr(t)+ (A_m-M^\top)\Delta r(t)\\
                     &= A_m \Delta r(t)+\Delta \theta(t)Mr(t),
  \end{split}
   \label{eq:errorsystem}
\end{align}
\begin{align}
   \Delta \dot{\theta}(t) = \dot{\hat{\theta}} (t) = -\gamma \Delta r^\top(t) P M r(t).
   \label{eq:theta_update}
\end{align}
% The dimension of the error system is three, with $\Delta r\in\mathbb{R}^2$ and $\Delta\theta\in\mathbb{R}$. 
% The time-varying bounded input $r(t)$ is presumed to be sinusoidal at an unknown frequency $\omega_0$ and an unknown phase $\phi$.

\begin{proposition}\label{prop1} For an appropriate choice of $P$ in the update law \eqref{eq:updatelaw}, the
  state estimates $\hat{r}(t)$ and parameter estimate
  $\hat{\theta}(t)$ exponentially converge to $r(t)$ and $\theta$
  respectively. That is the origin of the system formed by
  \eqref{eq:errorsystem} and \eqref{eq:theta_update} is globally
  exponentially stable.
\end{proposition}
% \begin{proposition} The error system \eqref{eq:errorsystem} is globally uniformly asymptotically stable in $\Delta r$.
% \end{proposition}
\begin{proof}
% The proof proceeds in several parts.  
Consider the candidate Lyapunov function
\begin{equation}
    V\big(\Delta r,\Delta\theta \big) = \frac{1}{2}\big(\Delta r^\top P\Delta r+\gamma^{-1} \Delta \theta^2 \big),
\label{eq:lyapunov}
\end{equation}
where $P\in\mathbb{R}^{2\times2}$ is the as yet unspecified positive definite matrix in the estimated frequency parameter update law \eqref{eq:updatelaw}.  The time derivative of $V\big(\Delta r, \Delta \theta \big)$, after substituting the error system \eqref{eq:errorsystem} and \eqref{eq:theta_update} and performing minor algebraic manipulations, is given by
\begin{equation}
\label{eq:Vdot}
\begin{aligned}
  \dot{V}\big(\Delta r(t), \Delta \theta(t) \big)
  % &=\Delta r^\top(t) P\Delta \dot{r}(t) + \gamma^{-1}   \Delta \theta(t) \Delta\dot{\theta}(t)\\
  % &=\Delta r^\top(t) P \Big( A_m \Delta r(t)+ \Delta \theta(t)Mr(t)\Big)\\
  % & \ \ \ - \Delta \theta(t)  \Delta r^\top(t) P M r(t)\\
  % &=\Delta r^\top(t) P A_m \Delta r(t) \\
  % & \ \ \ + \cancel{\Delta r^\top(t) P \Delta \theta(t)Mr(t)} \\
  % & \ \ \ - \cancel{\Delta \theta(t)  \Delta r^\top(t) P M r(t)}\\
  % &=\Delta r^\top(t) (P  A_m) \Delta r(t)\\\
  &=\frac{1}{2}\Delta r^\top(t) \big(A_m^\top P  + P  A_m \big) \Delta r(t).
\end{aligned}
\end{equation}
The matrix $A_m\in \mathbb{R}^{2\times2}$ is Hurwitz, thus given any positive definite symmetric matrix $Q\in\mathbb{R}^{2\times2}$, $P\in \mathbb{R}^{2\times2}$ is the unique positive-definite symmetric matrix that satisfies the linear Lyapunov equation 
\begin{equation}
\label{eq:symmetrix}
    -Q = {A_m}^\top P + PA_m.
\end{equation}
Thus, 

\begin{equation}
  \dot{V}\Big(\Delta r,  \Delta \theta \Big) = -\frac{1}{2}\Delta r^\top Q \Delta r \leq 0,
  \label{eq:Lyapunov_update}
\end{equation}
which is negative definite in $\Delta r(t)$ but only negative semi-definite in $\Delta r(t)$ and $\Delta \theta(t)$. Lastly, since $V$ is a quadratic function of $\Delta r, \Delta \theta$ we can apply a slightly different but global formulation of Theorem 4.8 of \cite{khalil2002nonlinear} where $V$ is radially unbounded to conclude that the error system is globally uniformly stable.

% globally and hence to encompass radially unbounded Lyapunov functions we conclude the system is , positive definite, and radially unbounded function of its arguments $\Delta r(t)$ and  $\Delta \theta(t)$. Given that $V$ bounded below by 0, and is non-increasing it is bounded above by its initial value, we can conclude that $\Delta r(t)$ and $\Delta \theta(t)$ are bounded.

To prove asymptotic stability of $\Delta r(t)$, we first show that\footnote{$L^p$ denotes the space of measurable functions with  $\|f\|_{p} < \infty$} $\Delta r(t) \in L^2 \cap L^{\infty}$. Let $\lambda_1>0$ be the smallest eigenvalue of $Q$, 
% (which is a positive definite matrix), 
it follows from the Rayleigh-Ritz theorem \cite{johnson1985matrix} that
\begin{align}
   0 \leq \Delta r^\top(t) \Delta r(t) \leq \frac{1}{\lambda_1} \Delta r^\top(t) Q \Delta r(t). 
\end{align}
From \eqref{eq:Lyapunov_update}, 
\begin{equation}
 \Delta r^\top Q \Delta r = -2 \dot{V}(\Delta r, \Delta \theta),
\end{equation}
thus 
\begin{align}
0 \leq \Delta r^\top(t) \Delta r(t) \leq -\frac{2}{\lambda_1}\dot{V}\big(\Delta r(t), \Delta \theta (t)\big). 
\end{align}
For
\begin{equation}
\bigg[\int_{0}^{\infty}\lVert \Delta r(\sigma)\lVert^2 d\sigma \bigg]^{1/2} =  \bigg[\int_{0}^{\infty}\Delta r^\top(\sigma) \Delta r(\sigma) d\sigma \bigg]^{1/2},
\end{equation}
then,
\begin{equation}
    \begin{aligned}
    & \bigg[\int_{0}^{\infty}\Delta r^\top(\sigma) \Delta r(\sigma) d\sigma \bigg]^{1/2} \\
    & \leq \bigg[\int_{0}^{\infty} -\frac{2}{\lambda_1} \dot{V}\big(\Delta r(\sigma), \Delta \theta (\sigma)\big) d\sigma \bigg]^{1/2} \\
    % &= \bigg[\int_{0}^{\infty} -\frac{2}{\lambda_1} d{V}\big(\Delta r(\sigma), \Delta \theta (\sigma)\big) \bigg]^{1/2}\\
    &= \bigg[-\frac{2}{\lambda_1} \bigg({V}\big(\Delta r(\infty), \Delta \theta (\infty)\big) - {V}\big(\Delta r(0), \Delta \theta (0)\big) \bigg) \bigg]^{1/2}\\ 
    & < \infty.
    \end{aligned}
\end{equation}
Thus, $\Delta r(t) \in L^2$. Also, since $\Delta r(t)$ is bounded for all $t \geq 0$, $\sup_{t \geq 0}|\Delta r(t)| < \infty,$
% \begin{equation}
%   \sup_{t \geq 0}|\Delta r(t)| < \infty,
% \end{equation}
thus $\Delta r(t) \in L^\infty$.
Finally, since $\Delta r(t), r(t)$, and $\Delta\theta(t)$ are all bounded, it follows from \eqref{eq:errorsystem} that $\Delta \dot{r}(t)$ is bounded and from \eqref{eq:updatelaw} that $\Delta \dot{\theta}(t)$ is bounded.

Thus, since $\Delta r(t) \in L^2 \cap L^\infty$, and $\Delta\dot r(t)$ is bounded, it  follows from Barbalat's Lemma (Lemma 2.12 and Corollary 2.9 in \cite{narendra1989stable}) that 
\begin{equation}
\lim_{t\to\infty} \Delta r(t)=0.
\label{eq:delta_r_0}
\end{equation}
%
% We observe from \eqref{eq:errorsystem} that $\Delta \dot{r}(t)$ is composed of sums and products of bounded functions hence $\Delta \dot{r}(t)$ is UC in time.
%
Furthermore, from \eqref{eq:errorsystem}, $\Delta \dot{r}(t)$ is the sums and products of bounded functions that are uniformly continuous
(UC) in time, thus it is UC and we conclude from Lemma 4.2 in \cite{slotine1991applied}, a variant of Barbalat's Lemma, that 
\begin{equation}
\lim_{t\rightarrow\infty} \Delta \dot{r}(t) = 0.
\end{equation}
Therefore, using the fact that the left-hand side of \eqref{eq:errorsystem} goes to zero, $A_m\Delta r(t) \to 0$ and $Mr(t) = [0 \quad \sin(\omega_0 t + \phi)]^{\top}$, we must have that:
\begin{equation}\label{eq:theta-converge}
  \lim_{t\to\infty}\Delta\theta(t)=0,
\end{equation}
i.e.\ $\hat{\theta}(t)$ converges to $\theta$. Since we showed before the error system is globally uniformly stable, we conclude from \eqref{eq:delta_r_0}, \eqref{eq:theta-converge} that it is also globally uniformly asymptotically stable.

Lastly, equations \eqref{eq:errorsystem} and \eqref{eq:theta_update} form a linear time-varying (LTV) system because $Mr(t) = [0 \quad \sin(\omega_0 t + \phi)]^{\top}$. Hence by a corollary of Theorem 4.11 in \cite{khalil2002nonlinear} any LTV system that is globally uniformly asymptotically stable must be globally exponentially stable.

\end{proof}

% \begin{corollary}\label{cor:exp-stability}
%     The origin of the system formed by \eqref{eq:errorsystem} and \eqref{eq:theta_update} is globally exponentially stable.
% \end{corollary}

% \begin{proof}
%     From the previous proposition the error system is globally uniformly asymptotically stable. Furthermore, equations \eqref{eq:errorsystem} and \eqref{eq:theta_update} form a linear time-varying (LTV) system since $Mr(t) = [0 \quad \sin(\omega_0 t)]^{\top}$. Therefore, by a corollary of Theorem 4.11 in \cite{khalil2002nonlinear} any LTV system that is globally uniformly asymptotically stable must be globally exponentially stable. \qed
% \end{proof}

\section{Online Internal Model Principle}
\label{sec:onlineIMP}
Given the stable adaptive estimator (\textbf{Problem} \ref{prob1}), we
now seek to understand if a closed-loop IMP-based system that relies
on the state and parameter estimates from said estimator is stable
(\textbf{Problem} \ref{prob2}).  To address, we rewrite the system in
\fref{fig2} and \eqref{eq:tf} as a 6-state interconnection.  Let the
$2^{\mathrm{nd}}$ order system in \eqref{eq:tf} be
$ \bm z_1\!\in\!\mathbb{R}^2$ with output $u_1\in\mathbb{R}$, the
time-varying version of damped oscillator in \eqref{eq:tf} be
$\bm z_2\!\in\!\mathbb{R}^2$ with output $u_2\in\mathbb{R}$, and the
delay block with a second-order Pad\'{e} approximation be
$\bm v\!\in\!\mathbb{R}^2$ with output $y\in\mathbb{R}$.  With sensory
error $e(t)=\hat r_1(t)-y(t)$, these subsystems can be written as
\begin{align}
\dot{\bm z}_1 &= 
\underbrace{\begin{bmatrix}0&1\\[2pt]0&-k_3\end{bmatrix}}_{A_1}\bm z_1
+\underbrace{\begin{bmatrix}0\\[2pt]1\end{bmatrix}}_{B_1} e,
\quad
u_1=\underbrace{\begin{bmatrix}k_2&k_1\end{bmatrix}}_{C_1}\bm z_1,\\[4pt]
\dot{\bm z}_2 &=
\underbrace{\begin{bmatrix}0&1\\[2pt]-\hat\omega^2(t)&-2\zeta\,\hat\omega(t)\end{bmatrix}}_{A_2(t)}\bm z_2
+\underbrace{\begin{bmatrix}0\\[2pt]1\end{bmatrix}}_{B_2} e,
\quad
u_2=\underbrace{\begin{bmatrix}k_4&0\end{bmatrix}}_{C_2}\bm z_2,\\[4pt]
\dot{\bm v} &=
\underbrace{\begin{bmatrix}0&1\\[2pt]-\tfrac{12}{\tau^2}&-\tfrac{6}{\tau}\end{bmatrix}}_{A_v}\bm v
+\underbrace{\begin{bmatrix}0\\[2pt]\tfrac{1}{\tau^2}\end{bmatrix}}_{D_v} u,
\quad
y=\underbrace{\begin{bmatrix}0&-12\,\tau\end{bmatrix}}_{C_v}\bm v+u,
\end{align}
where \(u:=u_1+u_2\).

Stack the states as
\[
x:=\begin{bmatrix}\bm z_1\\ \bm z_2\\ \bm v\end{bmatrix}
=\begin{bmatrix}z_{11}&z_{12}&z_{21}&z_{22}&v_1&v_2\end{bmatrix}^\top\in\mathbb{R}^6,
\]
and define the block matrices
\begin{equation*}
\begin{aligned}
& A(t):=\mathrm{blkdiag}\big(A_1,A_2(t),A_v\big),\quad
B:=\begin{bmatrix}B_1\\ B_2\\ \mathbf{0}_{2\times 1}\end{bmatrix}, \\
& D:=\begin{bmatrix}\mathbf{0}_{4\times 1}\\ D_v\end{bmatrix},
\end{aligned}
\end{equation*}

\begin{equation*}
\begin{aligned}
& F:=\begin{bmatrix}C_1 & \; C_2 & \; \mathbf{0}_{1\times 2}\end{bmatrix}
=\begin{bmatrix}k_2&k_1&k_4&0&0&0\end{bmatrix},\quad \\
& C:=\begin{bmatrix}C_1 & \; C_2 & \; C_v\end{bmatrix}
=\begin{bmatrix}k_2&k_1&k_4&0&0&-12\,\tau\end{bmatrix}.
\end{aligned}
\end{equation*}

With \(y=C x\), \(u=F x\), and \(e=\hat r_1-y\), the overall closed-loop system is given by
\begin{equation}\label{eq:ode-with-plant}
  \begin{aligned}
    \dot{x}(t)&=\big(A(t)+DF-BC\big) x(t)+B\,\hat r_1(t),
    \\
    y(t)&=C x(t).
  \end{aligned}
\end{equation}

Problem \ref{prob2} is then concerned with comparing $e(t)$ to the error generated if $x(t)$ had access to perfect information of $\omega_0$ and $r_1(t)$. Therefore, we define the closed-loop system with perfect information of the two aforementioned quantities by: 
\begin{equation}
    \dot{x}_c(t) = A_{c}x_c(t) + (DF - BC)x_c(t) + Br_1(t).
\end{equation}
where $A_{c} := \mathrm{blkdiag}(A_1,A_{\omega_0},A_{v}),$

\begin{equation}
% \begin{aligned}
A_{\omega_0} = \begin{bmatrix}0&1\\[2pt]- \omega_0^2&-2\zeta\,\omega_0\end{bmatrix},
% \end{aligned}
\end{equation}
and we seek to show that the origin is an exponentially stable point of the error system:
\begin{equation}\label{eq:imp-online-error}
    l(t) = x(t) - x_c(t).
\end{equation}

\begin{proposition}
    Assume that $k_1,k_2,k_3,k_4$ are chosen such that the matrix $A_{c} + DF - BC$ is Hurwitz and its eigenvalues have largest real part smaller than $-\lambda < 0$, then the error system \eqref{eq:imp-online-error} is globally exponentially stable.
\label{prop2}
\end{proposition}

\begin{proof}
First we notice from the adaptive design that $\Delta \hat{r}_1(t)$ is decoupled entirely. Further by Proposition \ref{prop1}, $\hat{r}_1(t) \to r_1(t) \text{ and } \hat{\omega}(t) \to \omega_0$ exponentially fast, therefore we can write:
\begin{equation}
% \begin{aligned}
    \hat{r}_1(t) = r_1(t) + c_0e^{-\nu_0 t}, \quad \hat{\omega}(t) = f(t) + \omega_0
    % & \hat{\omega}(t) = f(t) + \omega_0, \\
% \end{aligned}
\end{equation}
for some $c_0$ and $\nu_0 > 0$ and where\footnote{$\|\cdot\|$ refers to the $\ell_2$-norm} $\|f(t)\| \leq c_1 e^{-\nu_1 t}\|f(0)\|$ for some $\nu_1,c_1 > 0$. This implies we can decompose $A(t) = A_{c} + A_{p}(t)$, where $A_{p}(t) := \mathrm{blkdiag}(\mathbf{0}_{2 \times 2},A_{p_2},\mathbf{0}_{2 \times 2}),$
\begin{equation}
% \begin{aligned}
% & A_{p}(t) := \mathrm{blkdiag}(\mathbf{0}_{2 \times 2},A_{p_2},\mathbf{0}_{2 \times 2}), \\
A_{p_2} = \begin{bmatrix}0&0\\[2pt] - f(t)^2 - 2f(t)\,\omega_0& -2 \zeta\,f(t)\end{bmatrix}.
% \end{aligned}
\end{equation}
Since $f(t)$ decays exponentially with time we also know $\|A_p(t)\| \leq c_2e^{-\nu_1 t} \|A_p(0)\|$ for some $c_2> 0$.
Now rewriting \eqref{eq:ode-with-plant} with the above we get
\begin{multline}
   \dot{x}(t)  = A_{c}x(t) + A_{p}(t)x(t) + (DF - BC)x(t) + \\
   Br_1(t) + Bc_0e^{-\nu_0 t},
\end{multline}
and the time derivative of \eqref{eq:imp-online-error} is given by
\begin{equation}
\begin{aligned}
\dot{l}(t) = (A_{c} + DF - BC)l(t) + A_p(t)x(t) + Bc_0e^{-\nu_0 t}.
\end{aligned}
\end{equation}
Hence exponential stability of the above boils down to asserting the convergence of a system
\begin{equation}\label{eq:perturbation}
    \dot{\xi}(t) = H\xi(t) + \delta(t),
\end{equation}
where $H = A_{c} + DF - BC$ is Hurwitz and $\delta(t) = A_{p}(t)x(t) + Bc_0e^{-\nu_0 t}$.
To do so we show first that $x(t)$ is uniformly bounded above. Given that $A_{c} + DF - BC$ is Hurwitz and $A_p(t)$ vanishes exponentially we know by Lemma 2.2 of \cite{narendra1989stable} that the origin of \eqref{eq:ode-with-plant} without input $Br_1(t) + Bc_0e^{-\nu_0 t}$ 
% \textcolor{red}{MS: you are using $\lambda$ for 2 different things: the perturbation and $F$. I think this should be $Br_1(t) + BC_oe^{-\nu_o t}$}
is globally exponentially stable. Therefore by Theorem 4.11 of \cite{khalil2002nonlinear} the fundamental matrix of \eqref{eq:ode-with-plant} without input satisfies 
\begin{equation}\label{eq:state-transition-exp-decay}
    \|\Phi(t,0)\|\leq \kappa e^{-K t}, \qquad \forall t \geq 0,
\end{equation}
for some $\kappa,K > 0$. Hence a variation of constants formula applied to \eqref{eq:ode-with-plant} with the input yields
\begin{multline}
   \|x(t)\| \leq \|\Phi(t,0)\|\|x(0)\| \\
   + \int_{0}^{t} \|\Phi(t,\tau)\|\, \|Br_1(t) + Bc_0e^{-\nu_0 t}\|\, d\tau,
\end{multline}
which is clearly bounded above by a constant $c_3$ since $Br_1(t) + Bc_0e^{-\nu_0 t}$ is bounded above and \eqref{eq:state-transition-exp-decay} holds. Therefore $\|x(t)\|$ is bounded above and
\begin{equation}
    \|\delta(t)\| \leq \|A_{p}(t)\|\|x(t)\| + \|Bc_0e^{-\nu_0 t}\| \leq c_4e^{-\min(\nu_1,\nu_0)t}
\end{equation}
for some $c_4 > 0$.
Applying now the variation of constants formula to \eqref{eq:perturbation} and letting $\nu = \min(\nu_1,\nu_0)$, yields
\begin{equation}
\begin{aligned}
   \|\xi(t)\| &\leq \|e^{Ht}\|\|\xi(0)\| 
    + \int_{0}^{t}\|e^{H(t-\tau)}\|\|\delta(\tau)\|d\tau \\ 
    % &  \leq c_5e^{-\lambda t}\|\xi(0)\| 
    % + \int_{0}^{t} c_5e^{-\lambda(t-\tau)}c_4e^{-\nu \tau}d\tau \\
    & \leq c_5e^{-\lambda t}\|\xi(0)\| 
    + c_5c_4e^{-\lambda t}\int_{0}^{t} e^{(\lambda - \nu)\tau}d\tau.
\end{aligned}
\end{equation}
If $\lambda = \nu$, $\xi(t)$ is clearly exponentially decaying. If $\lambda \neq \nu$ the above equals
\begin{equation}
\begin{aligned}
    & c_5e^{-\lambda t}\|\xi(0)\| + c_5c_4e^{-\lambda t}\left(\frac{e^{(\lambda - \nu)t}- 1}{\lambda - \nu}\right),
    % & c_5e^{-\lambda t}\|\xi(0)\| + c_5c_4
    % \left(\frac{e^{-\nu t} - e^{-\lambda t} }{\lambda - \nu}\right)
\end{aligned}
\end{equation}
which is exponentially decaying with rate $\min(\lambda,\nu)$. 
\end{proof}

Lastly since $x(t) \to x_c(t), \hat{r}_1(t) \to r_1(t)$ exponentially fast, we know that the error $e(t) \to e_c(t) = r_1(t) - Cx_c(t)$ exponentially fast as well. This answers \textbf{Problem} \ref{prob2} and justifies the use of the online IMP-based architecture.

\section{Experiments and Simulations}
All \emph{Eigenmannia virescens} were obtained from company vendors and were housed following published guidelines \cite{hitschfeld2009effects}. All experimental procedures in this paper were approved by the Johns Hopkins Animal Care and Use Committee and were in compliance with guidelines established by the National Research Council and the Society for Neuroscience. The fish refuge tracking experiments were conducted using a protocol similar to previous work
\cite{roth2011stimulus}.

\begin{figure}[h]
    \centering
\includegraphics[width=0.96\linewidth]{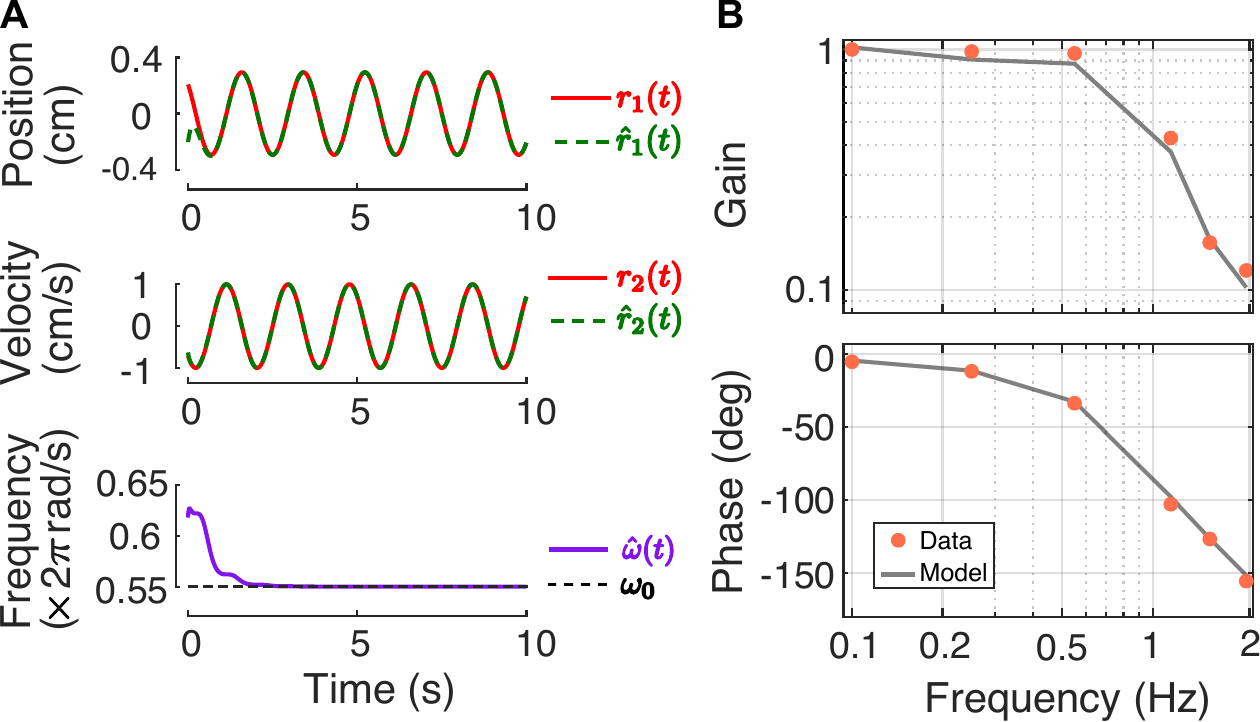}
    \caption{Simulated results corroborate the analytical proof and experimental results. (A) Top and middle: the identified refuge position $\hat{r}_1(t)$ and velocity $\hat{r}_2(t)$ (green dashed) converge to the actual refuge position $r_1(t)$ and velocity $r_2(t)$ (red). Bottom: the identified stimulus frequency $\hat{\omega}(t)$ converges to the actual stimulus frequency $\omega_0$. (B) Comparison between the Bode plot of experimental results (data) from six sinusoidal frequencies in orange dots vs simulated results from the adaptive-IMP modeling illustrated by grey curves.}
    \label{fig3}
\end{figure}

The analytical proof shows that given the external single sine input
$r(t) = [r_1(t) \ r_2(t)] ^\top$ with frequency $\omega_0$, the
identified refuge position $\hat{r}_1(t)$, refuge velocity
$\hat{r}_2(t)$, and frequency $\hat{\omega}(t)$ exponentially converge
to $r_1(t)$, $r_2(t)$, and $\omega_0$ despite the initial discrepancy
at $t = 0$. Here, we simulate the adaptive identifier in
Section~\ref{sec:adaptive_model} with various parameter choices in
MATLAB as illlustrative examples. An example with a stimulus frequency
of 0.55 Hz (\fref{fig3}A) demonstrates that the estimated frequency
converges to 0.55 Hz in about 2.5 seconds.  By tuning the parameters
of the adaptive identifier, we are able to adjust the speed of
adaptive identification. Both analytical proofs and simulations
suggest that our proposed adaptive identifier can model how fish
adaptively identify the frequency of a single sine stimulus over time
during refuge tracking.

We simulated the cascade of the adaptive identification, time-varying
controller, and time-invariant plant and delay together with
the state-space form of the closed-loop system in \fref{fig2} and
Sections \ref{sec:adaptive_model} and \ref{sec:onlineIMP}. We fit the
model parameters to newly collected experimental data. The parameters $k_1$, $k_2$, and $k_3$, and
$\tau$ are constant across frequencies while $k_4$ and $\zeta$ are
fitted differently for different single sine stimuli to capture the
general trend in the frequency response gain and phase for the
averaged Bode plot in fish single-sine tracking (\fref{fig3}B).
The parameters used here make $A_{c} + DF - BC$ Hurwitz, thus satisfying the stability assumption in \textbf{Proposition} \ref{prop2}. 

\section{Conclusion}

In this paper, we introduce a candidate model for how fish track predictable stimuli such as single sine wave.  We propose that fish adaptively identify the single sine stimulus and its frequency. The identified frequency is embedded into an IMP-based controller to achieve reference tracking. Using perturbation arguments, we prove that the closed-loop system with internal estimates of frequency and state converges exponentially to the closed-loop system with perfect information. Numerical simulations illustrate the analytical proofs and fit the frequency response of \emph{Eigenmannia} well in the single sine refuge tracking task, offering evidence for the biological relevance of the model. 

Although the present model provides a rigorous framework for the task
of tracking a single sine stimulus, we make no claims as to how it is
achieved mechanistically in the fish brain. Furthermore, our model
restricts the input stimulus to single sinusoids, which allows us to
recast the adaptive identifier as a linear time-varying system and
prove exponential convergence of the IMP-based closed-loop
system. However, for other types of stimulus, such as sum-of-sines
\cite{roth2011stimulus}, the system is nonlinear. Future work seeks to
explore the neurological basis of stimulus adaptation and extend the
model to explain how fish manage to track a more general class of
input signals, including sum-of-sines \cite{roth2011stimulus} and noise\cite{yang2020comparison}. We hope that recent advances in Loewner-based system identification will provide promising tools for extracting accurate low-order models of fish tracking data \cite{HONARPISHEH2024199}, while nonlinear adjustment strategies reminiscent of phase-locked loop mechanisms or machine learning techniques like recurrent neural networks \cite{sussillo2009generating} will help modeling the nonlinear behavior.
% We hope that nonlinear adjustment
% strategies reminiscent of phase-lock loop mechanisms or modern machine
% learning techniques like recurrent neural networks
% \cite{sussillo2009generating} will be particularly useful. 

\section{Acknowledgment}
We thank Dr. Michael G.T. Wilkinson for his pilot work on this project
and Dr. Jinwoo Choi for his help with experimental data
collection. The authors wrote all text. AI was used minimally to identify typos and make minor wording suggestions.

\bibliographystyle{IEEEtran} % We choose the "plain" reference style
\bibliography{reference}

\end{document}